\documentclass[aps, prev,preprintnumbers,floatfix,nofootinbib, notitlepage, twocolumn,superscriptaddress, prd]{revtex4-1}

\pdfoutput=1

\usepackage{mathrsfs, amsmath, amsthm, amssymb, color, epstopdf, verbatim, enumerate, graphicx, hyperref}

\newcommand{\betaterm}{\frac{\beta}{\beta+1}}

\newcommand{\veff}{V_{\text{eff}}}

\newcommand{\ql}{Q_{\text{L}}}

\begin{document}

\title{Dark photon constraints from a 7.139 GHz cavity haloscope experiment}

\author{Dong He}
\affiliation{Key Laboratory of Low-Dimensional Quantum Structures and Quantum Control of Ministry of Education, Key Laboratory for Matter Microstructure and Function of Hunan Province, Department of Physics and Synergetic Innovation Center for Quantum Effects and Applications, Hunan Normal University, Changsha 410081, China}
\author{Jie Fan}
\affiliation{Institute of Physics, Chinese Academy of Sciences, Beijing, 100190, China}
\author{Xin Gao}
\affiliation{College of Physics, Sichuan University, Chengdu 610065, China}
\author{Yu Gao}
\affiliation{Key Laboratory of Particle Astrophysics, Institute of High Energy Physics, Chinese Academy of Sciences, Beijing 100049, China}
\author{Nick Houston}
\affiliation{Institute of Theoretical Physics, Faculty of Science, Beijing University of Technology, Beijing 100124, China}
\author{Zhongqing Ji}
\affiliation{Institute of Theoretical Physics, Faculty of Science, Beijing University of Technology, Beijing 100124, China}
\author{Yirong Jin}
\affiliation{Beijing Academy of Quantum Information Sciences, Beijing 100193, China}
\author{Chuang Li}
\affiliation{College of Mechanical and Electrical Engineering, Wuyi University, Nanping 354300, China}
\author{Jinmian Li}
\affiliation{College of Physics, Sichuan University, Chengdu 610065, China}
\author{Tianjun Li}
\affiliation{CAS Key Laboratory of Theoretical Physics, Institute of Theoretical Physics, Chinese Academy of Sciences, Beijing 100190, China}
\affiliation{School of Physical Sciences, University of Chinese Academy of Sciences, No. 19A Yuquan Road, Beijing 100049, China}
\author{Shi-hang Liu}
\affiliation{Key Laboratory of Low-Dimensional Quantum Structures and Quantum Control of Ministry of Education, Key Laboratory for Matter Microstructure and Function of Hunan Province, Department of Physics and Synergetic Innovation Center for Quantum Effects and Applications, Hunan Normal University, Changsha 410081, China}
\author{Jia-Shu Niu}
\affiliation{Institute of Theoretical Physics, Shanxi University, Taiyuan, 030006, China}
\author{Zhihui Peng}
\affiliation{Key Laboratory of Low-Dimensional Quantum Structures and Quantum Control of Ministry of Education, Key Laboratory for Matter Microstructure and Function of Hunan Province, Department of Physics and Synergetic Innovation Center for Quantum Effects and Applications, Hunan Normal University, Changsha 410081, China}
\author{Liang Sun}
\affiliation{Institute of Physics, Chinese Academy of Sciences, Beijing, 100190, China}
\author{Zheng Sun}
\affiliation{College of Physics, Sichuan University, Chengdu 610065, China}
\author{Jia Wang}
\affiliation{Institute of Physics, Chinese Academy of Sciences, Beijing, 100190, China}
\author{Puxian Wei}
\affiliation{College of Physics and Optoelectronic Engineering,
Department of Physics, Jinan University, Guangzhou 510632, China}
\author{Lina Wu}
\affiliation{School of Sciences, Xi'an Technological University, Xi'an 710021, P. R. China}
\author{Zhongchen Xiang}
\affiliation{Institute of Physics, Chinese Academy of Sciences, Beijing, 100190, China}
\author{Qiaoli Yang}
\affiliation{College of Physics and Optoelectronic Engineering,
Department of Physics, Jinan University, Guangzhou 510632, China}
\author{Chi Zhang}
\affiliation{Institute of Physics, Chinese Academy of Sciences, Beijing, 100190, China}
\author{Wenxing Zhang}
\affiliation{Tsung-Dao Lee Institute and School of Physics and Astronomy, Shanghai Jiao Tong University, 800 Dongchuan Road, Shanghai 200240, China}
\author{Xin Zhang}
\affiliation{National Astronomical Observatories, Chinese Academy of Sciences, 20A, Datun Road, Chaoyang District, Beijing 100101, China}
\affiliation{School of Astronomy and Space Science, University of Chinese Academy of Sciences, Beijing 100049, China}
\author{Dongning Zheng}
\affiliation{Institute of Physics, Chinese Academy of Sciences, Beijing, 100190, China}
\author{Ruifeng Zheng}
\affiliation{College of Physics and Optoelectronic Engineering,
Department of Physics, Jinan University, Guangzhou 510632, China}
\author{Jian-yong Zhou}
\affiliation{Key Laboratory of Low-Dimensional Quantum Structures and Quantum Control of Ministry of Education, Key Laboratory for Matter Microstructure and Function of Hunan Province, Department of Physics and Synergetic Innovation Center for Quantum Effects and Applications, Hunan Normal University, Changsha 410081, China}
\medskip

\begin{abstract}
The dark photon is a promising candidate for the dark matter which comprises most of the matter in our visible Universe. 
Via kinetic mixing with the Standard Model it can also be resonantly converted to photons in an electromagnetic cavity, offering novel experimental possibilities for the discovery and study of dark matter. 
We report the results of a pathfinder dark photon dark matter cavity search experiment performed at Hunan Normal University and the Institute of Physics, Chinese Academy of Sciences, representing the first stage of the APEX (Axion and dark Photon EXperiment) program.
Finding no statistically significant excess, we place an upper limit on the kinetic mixing parameter $|\chi|<3.7\times 10^{-13}$ around $m_A\simeq 29.5$ $\mu$eV at 90\% confidence level.
This result exceeds other constraints on dark photon dark matter in this frequency range by roughly an order of magnitude.
\end{abstract}

\collaboration{APEX Collaboration}

\maketitle{}

{\bf Introduction.}
Overwhelming evidence exists that a large majority of the matter in our Universe is `dark', in that it interacts very feebly or not at all with the Standard Model (SM)~\cite{Rubin:1982kyu, Begeman:1991iy, Taylor:1998uk, Natarajan:2017sbo, Markevitch:2003at, Planck:2018vyg}.
In the $\Lambda$CDM model, this dark matter (DM) is weakly interacting, non-relativistic, and cosmologically stable. 
Beyond this point not much is known about the underlying nature of DM, or its interactions with SM particles.

In light of this, the dark photon is a promising DM candidate. 
As a spin-1 gauge boson associated to some additional U(1) symmetry, it is firstly one of the simplest possible extensions to the SM~\cite{Essig:2013lka, Ghosh:2021ard, Caputo:2021eaa}. 
Having the same quantum numbers as a SM photon, dark photons can also convert into SM photons via kinetic mixing~\cite{Holdom:1986eq, Holdom:1985ag}, as described by the Lagrangian
\begin{equation}
  \mathcal{L} = -\frac{1}{4}(F^{\mu \nu}F_{\mu \nu} +F_d^{\mu \nu}F_{d\mu \nu} - 2\chi F^{\mu \nu}F_{d\mu \nu} - 2 m_A^{2} A_d^{2}),
  \label{eq: lagrangian}
\end{equation}
where $F^{\mu \nu}$ and $F_d^{\mu \nu}$ are the electromagnetic and dark photon field strength tensors respectively, $\chi$ is the kinetic mixing parameter, $m_A$ is the dark photon mass, and $A_d^\mu$ is the dark photon gauge field. 
If $m_A$ and $\chi$ are both sufficiently small, then the dark photon should be stable on cosmological timescales~\cite{Pospelov:2008jk}, and hence an attractive DM candidate. 

Sufficiently light dark photons are best described as a coherent wave oscillating at a frequency set by $m_A$, rather than a collection of distinct particles, leading to interesting phenomenology and experimental possibilities.
The degree of coherence here is controlled by the DM velocity distribution, and in particular $v_\mathrm{DM}^2\sim 10^{-6}$~\cite{Turner:1990qx, Jimenez:2002vy}.

Several known mechanisms are capable of producing a relic population of dark photons, such as displacement of the dark photon field via quantum fluctuations during inflation. 
These fluctuations provide the initial displacement for dark photon field oscillations, which commence once Hubble friction becomes ineffective ~\cite{Graham:2015rva}. 
Other mechanisms are also possible, as described in~\cite{Caputo:2021eaa, Arias:2012az}.

This type of DM can then be detected via kinetic mixing: when dark photons convert into SM photons inside a high $Q$ electromagnetic cavity, they create a weak electromagnetic signal inside the cavity which can be detected by a sensitive receiver chain. 
This type of detector is typically called a haloscope and was originally developed to search for axion DM~\cite{Sikivie:1983ip}, although in recent years it has also become a powerful tool search for dark photon DM as well~\cite{Brubaker:2016ktl, ADMX:2018gho, Nguyen:2019xuh, HAYSTAC:2020kwv, CAPP:2020utb, Cervantes:2022epl, Cervantes:2022gtv, McAllister:2022ibe, Cervantes:2022yzp, Schneemann:2023bqc, Tang:2023oid}. 
In natural units ($\hbar = c = 1$) the SM photon frequency $f$ is connected to the dark photon energy $E_{d}$ via the resonance condition $2\pi f = E_{d} \simeq m_A$.

The resulting dark photon signal power is~\cite{Cervantes:2022epl} 
\begin{equation}
    P_{s} = P_0 \betaterm L(f, f_0, \ql)\,,\quad
    P_{0} = \eta\chi^2 m_A \rho\veff \ql\,,
    \label{eq: power}
\end{equation} 
where $\beta$ is the cavity coupling coefficient, $\eta$ is a signal attenuation factor, $\rho\simeq0.45\,\rm{GeV}/\rm{cm}^3$ is the local DM density, and $\ql$ is the cavity loaded quality factor.
The Lorentzian term 
 \begin{equation} 
    L(f, f_0, \ql) = 1/(1+4\Delta^2)\,,\quad \Delta \equiv \ql (f-f_0)/f_0\,,
    \label{eq: lorentzian}
\end{equation}
is a detuning factor dependent on $\ql$, the SM photon frequency $f$ and the cavity resonance frequency $f_0$. 
The effective volume of the cavity is meanwhile given by
\begin{equation}
    V_{\rm eff} = \frac{\left (\int dV {\bf E}(\vec{x}) \cdot {\bf A_d}(\vec{x})\right )^2}{\int dV |{\bf E}(\vec{x})|^2|{\bf A_d}(\vec{x})|^2}\,,
    \label{eq: veff}
\end{equation}
which can be understood as the overlap between the dark photon and the corresponding induced electric field ${\bf E}({\vec{x}})$. 
We assume that the physical cavity size is much smaller than the dark photon de Broglie wavelength.

Haloscope experiments typically search for DM as a narrow spectral power excess above a thermal noise floor.
The noise power $P_n$ arises from the blackbody radiation of the cavity itself, along with added Johnson noise from the receiver chain used to extract signal power from the cavity.
In the Rayleigh-Jeans limit (where $k_B T_{n} >> hf$), we have 
\begin{equation}
    P_n \simeq G k_B b T_{n}\,,
    \label{eq: noise power}
\end{equation} 
where $k_B$ and $h$ are the Boltzmann and Planck constants respectively, $G$ is the system gain, $b$ is the frequency bin width, and $T_{n}$ is the noise temperature. 

In practice the measured $P_n$ is typically the average of $N = b \Delta t$ power spectra, where $N$ is a large number, and $\Delta t$ is the integration time.
The corresponding SNR is then $P_s/\sigma_{P_n} = (P_s/P_n)\sqrt{b \Delta t}$, where $\sigma_{P_n} \simeq P_n/\sqrt{N}$.

\begin{figure}[t]
	\centering
 	\includegraphics[width=0.8\columnwidth]{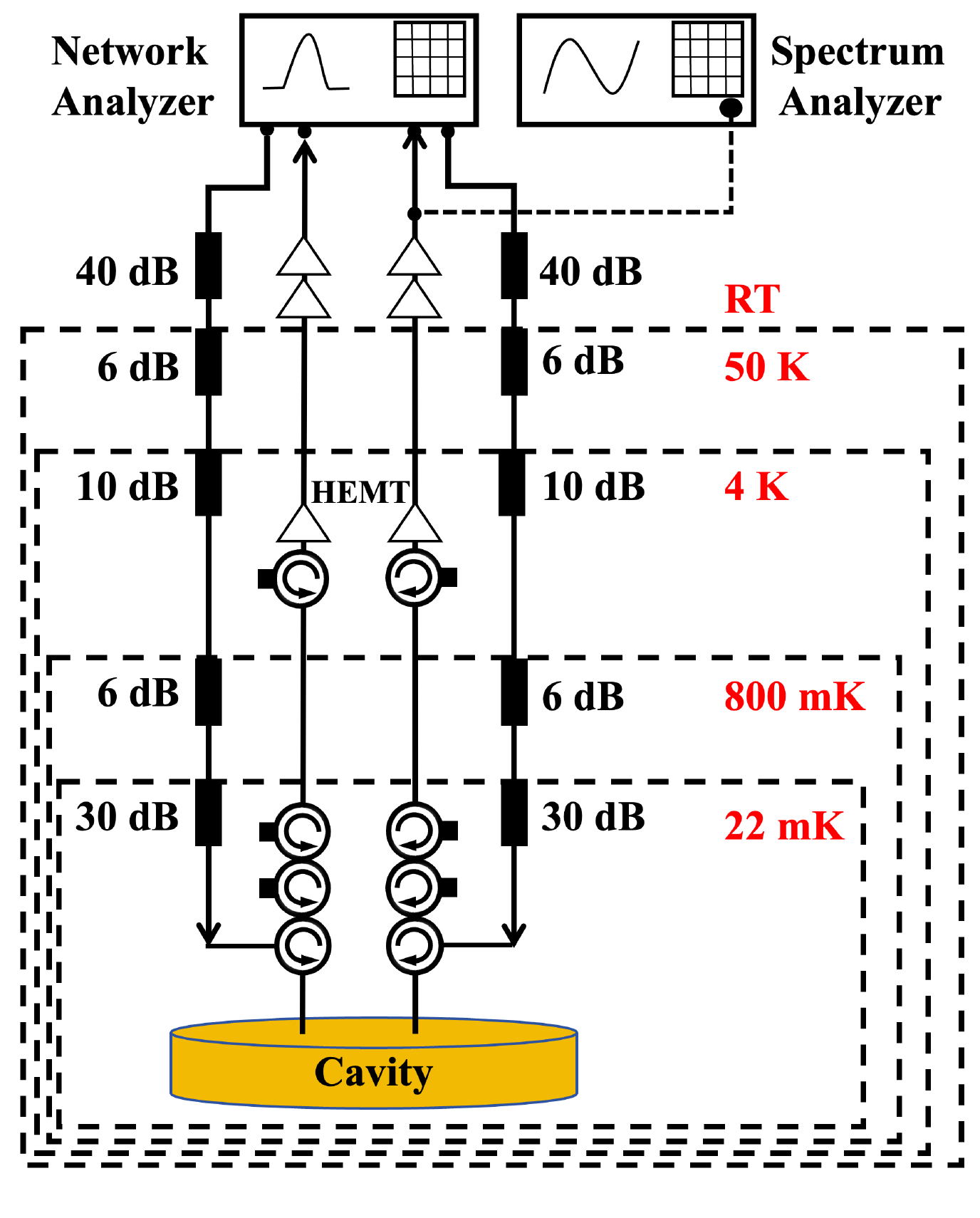}
	\caption{
        Experimental diagram.
        Different boxes represent different temperature layers, with the cavity located in the coldest region.
        The four-port vector network analyzer measurement consists of attenuators for each temperature layer, three circulator, an isolator, a HEMT, and two room-temperature amplifiers.
        A spectrum analyzer is used to measure spectrum power.
    }
	\label{fig: experimental diagram}
\end{figure}

In this paper, we describe a cavity haloscope experiment based at Hunan Normal University and the Institute of Physics, Chinese Academy of Sciences, designed to serve as a pathfinder for the APEX (Axion and dark Photon EXperiment) program.
We present the underlying experimental methodology, followed by an analysis of a dataset taken in 2023.
From this analysis, we place an upper limit on the dark photon kinetic mixing parameter $|\chi|<3.7\times 10^{-13}$ around $m\simeq 29.5\mu$eV to 90 \% confidence level, which exceeds other constraints on dark photon dark matter in this frequency range by roughly an order of magnitude.
Discussion and conclusions are presented in closing.

\textbf{Experimental methodology.}
As indicated in Fig. ~\ref{fig: experimental diagram}, measurements are carried out in a Bluefors LD 400 dilution refrigerator with a base temperature of about $22\, \mathrm{mK}$. 
The signal from the cavity is first amplified by a cryogenic HEMT amplifier (ZW-LNA2.4-9A). 
At around 7.139 GHz, the amplifier noise temperature is 5.5 K and the gain is about 36 dB, per manufacturer calibration data. 

Between the HEMT and the cavity, there are four cryogenic microwave circulators (CIRG00811A005).
Three circulators are terminated with $50\,\Omega$ terminators and work as isolators.
These isolators and circulators prevent the HEMT amplifiers from injecting noise into the cavity. 
The nominal total losses between the cavity and the HEMT are around 3 dB, which corresponds to an $\eta$ of about 0.5.

The signal is further amplified at room temperature using two WQF0118-30-15 amplifiers, each with a gain of 36 dB. 
The signal is then injected into the appropriate measurement equipment. 
The main measuring equipments used in our experiment are a spectrum analyzer (Keysight N9020B) and a network analyzer (Keysight N5231B).

We measure the transmission spectrum and reflection spectrum through the two ports of this cavity by using a four-port network analyzer. 
The total attenuation due to cables in the low and room temperature sections is about 20 dB, in the room temperature stage we also mount two 20 dB attenuators in each of the two output ports of the network analyzer.
The scanning frequency range is set to 15 MHz, and the probe power reaching the cavity is about $-142\  \mathrm{dBm}$\ which corresponds to an average photon number $\langle n\rangle=P / 2\hbar \omega_r \kappa  $\ inside the cavity of less than 1.
Taking the maximum value of the measured reflection data as background, we normalize the reflection spectrum and transmission spectrum.
We find that the cavity has a dissipation rate $\kappa= 2\pi\times0.6 \ \mathrm{MHz}$, a $Q_L$ value of 11006, reflection and transmission coefficients of $R^2 = 0.758$ and $T^2 = 0.175$, and a coupling $\beta = 0.9539$ at $7.139\  \mathrm{GHz}$.

For the dark photon search, emission power from the cavity is measured using the Keysight N9020B Spectrum Analyzer.
In this case, we set the frequency range of our scan to $ 7.1389 \ \mathrm{GHz} \pm 75\ \mathrm{ kHz}$, with a step size between points of 750 Hz. 
Each frequency point is scanned $10^5$ times continuously. We repeated this procedure $10^3$ times to give $10^8$ measurements per frequency point. 
The scan time is set to automatic: given our chosen parameters, the time to automatically scan $10^5$ times is 22.1 ms. 
By repeating this $10^3$ times, we arrive at a total integration time per point of 22.1 s.

The effective volume $V_{\rm eff}$ cannot be directly measured, and so needs to be computed via simulation.
In practice we expect the dark photon field to be spatially uniform over scales much larger than the cavity size, so that the field orientation can be specified via a constant unit vector $\hat n$.
With ${\bf A_d(x)}\propto {\bf \hat n}$, Eq.~\eqref{eq: veff} then becomes
\begin{equation}
    \veff = \frac{\left(\int dV {\bf E}(\vec{x})\cdot {\bf\hat n}\right)^2}
    {\left (\int dV|{\bf E}(\vec{x})|^2 \right )} 
    =\frac{\left(\int dV {\bf E}(\vec{x})\right)^2}
    {\left (\int dV|{\bf E}(\vec{x})|^2 \right )}
    \langle \cos^2\theta \rangle_T\,,
\end{equation} 
where $\theta$ is the (unknown) angle between ${\bf E(x)}$ and ${\bf A_d(x)}$ during a typical measurement integration time. 
Assuming the dark photon is randomly polarized, $\langle \cos^2\theta \rangle_T =1/3$ for a cavity haloscope of this type \cite{Arias:2012az}.

\begin{figure}[h]
	\centering
	\includegraphics[width=0.8\columnwidth]{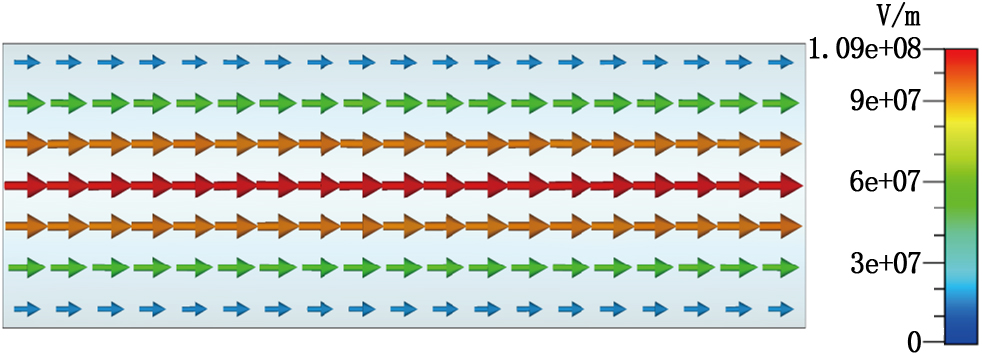}    
	\caption{Electric field distribution of the $\rm TM_{010}$ cavity mode, simulated for a cylindrical cavity of radius 16.2 mm, and height 90 mm.}
	\label{fig: electric field distribution}
\end{figure}

The electric field distribution of the cavity operating in the $\rm TM_{010}$ mode was simulated using CST Microwave Studio, as illustrated in Fig.~\ref{fig: electric field distribution}. 
The simulated $V_{\rm eff} \langle \cos^2\theta \rangle^{-1}_T $ was $51.3\, \mathrm{cm^3}$, with a cylindrical cavity radius of 16.2 mm, and height of 90 mm.
The electromagnetic field distribution was obtained from the intrinsic mode simulation, so no input/output ports were added.
However, simulations with an excitation port were also performed; it was found in this case that the resonance frequency increased and the loaded quality factor decreased, meanwhile other aspects of the cavity performance were not significantly impacted.





The relevant experimental parameters are summarised in Table \ref{tab: experimental parameters}.
\begin{table}[h]
    \centering
    \begin{tabular}{|l|l|l|l|l|l|l|l|l|}
    \hline
    $\beta$ & $f_0$  & $Q_L$ & $V_{\rm eff}$ & $G$ & $\eta$  &$b$ & $t_{\rm int}$      \\ \hline
    0.9539  & 7.139 {\rm GHz}  & 11006   &17.1 {\rm ml} & 88 {\rm dB} & 0.5 & 20 {\rm Hz} & 22.1 {\rm s} \\
    \hline
    \end{tabular}
    \caption{Key experimental parameters.}
    \label{tab: experimental parameters}
\end{table}

\textbf{Data analysis.}
To analyse the resulting data we broadly follow the prescription outlined in other dark photon cavity haloscope searches, such as Ref.~\cite{Cervantes:2022epl}.
That said, as a pathfinder experiment our approach is simpler in that we do not tune the cavity, and analyse only a single power spectrum.

This being the case, various steps in the standard analysis prescription are not required, such as the combination of various power spectra measured at different resonant frequencies.
Furthermore, the frequency range under consideration is sufficiently narrow that gain frequency dependence is also negligible, so we do not need to employ e.g. the Savitzky-Golay filtering common to other haloscope data analyses.

\begin{figure}[h]
	\centering
	\includegraphics[width=0.8\columnwidth]{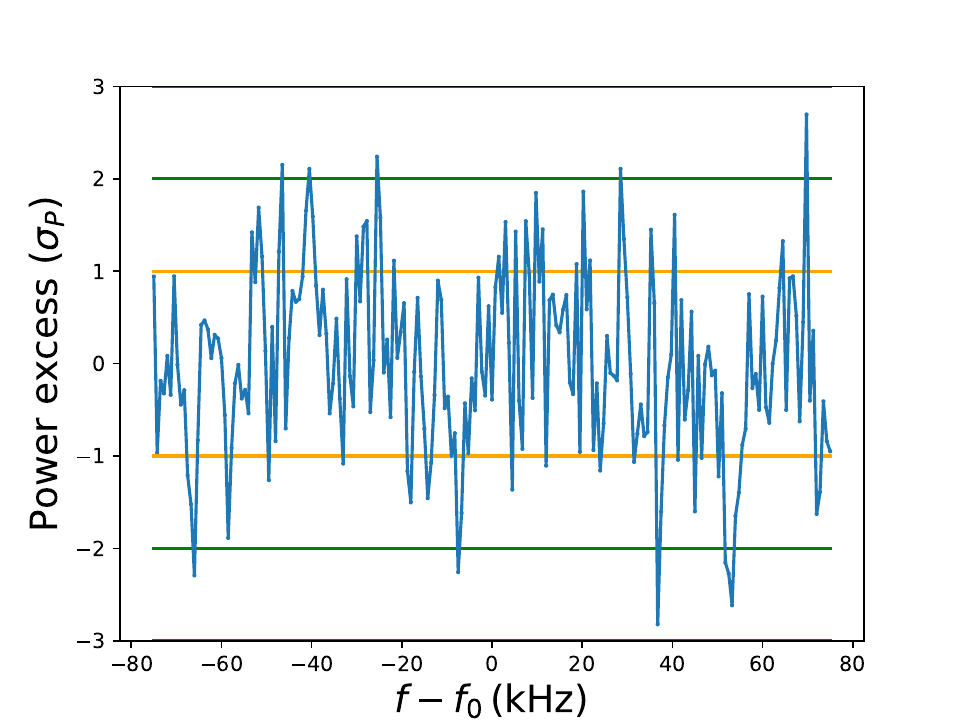}
	\caption{Spectral power excess, in units of the standard deviation.
    As can be seen, all data points lie within 3 standard deviations of the mean.
    }
	\label{fig: power excess}
\end{figure}

Subtracting firstly the mean power $\left\langle P\right\rangle\simeq 3.5\times10^{-21}$W from the measured power spectrum power we find the power excess $P_e = P - \left\langle P\right\rangle$, shown in Fig.~\ref{fig: power excess} in units of the corresponding standard deviation $\sigma_{P}$.
We can see that the data are compatible with the null hypothesis.

\begin{figure*}[t]
	\centering
	\includegraphics[width=0.95\columnwidth]{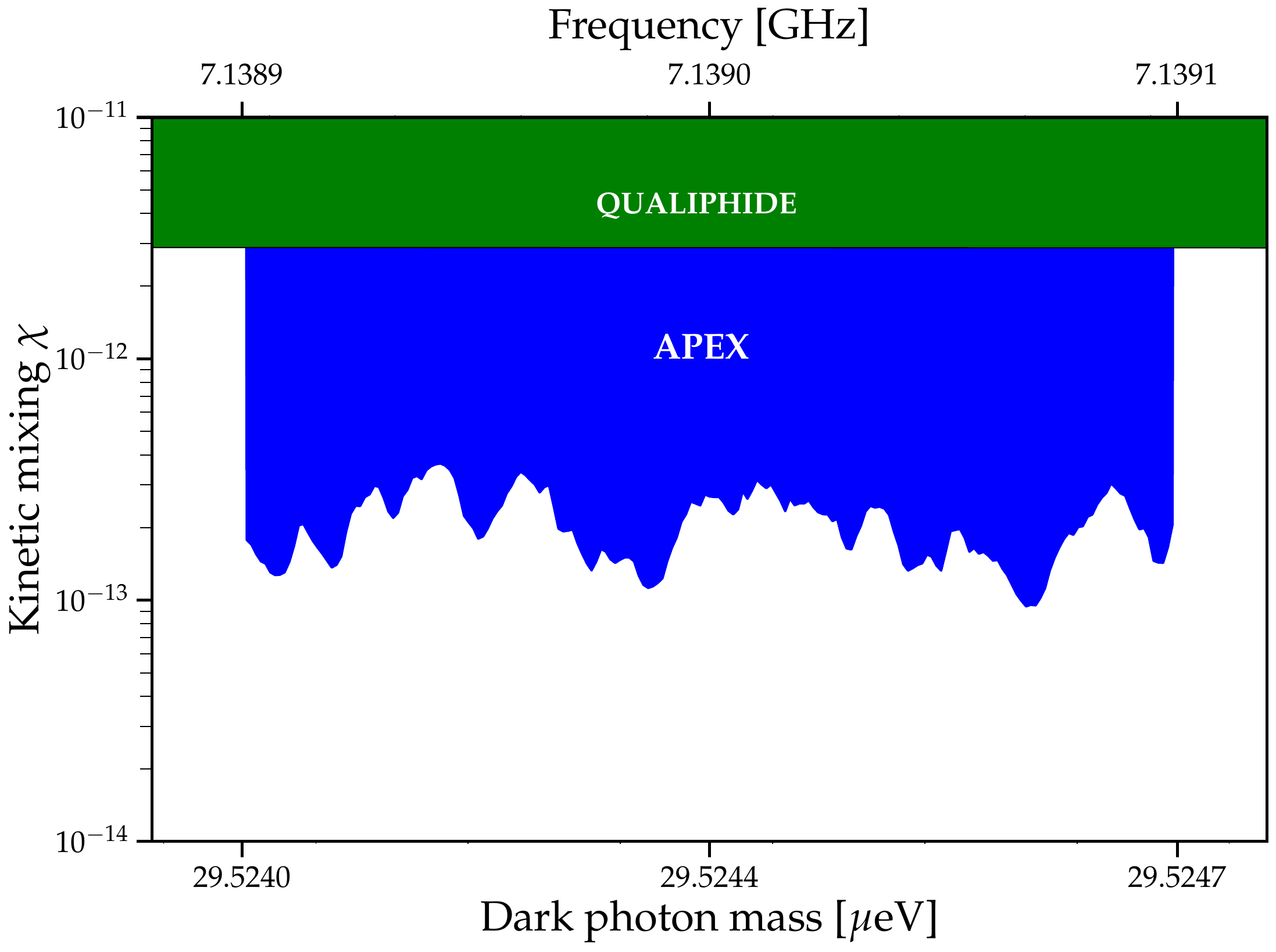}
	\includegraphics[width=0.87\columnwidth]{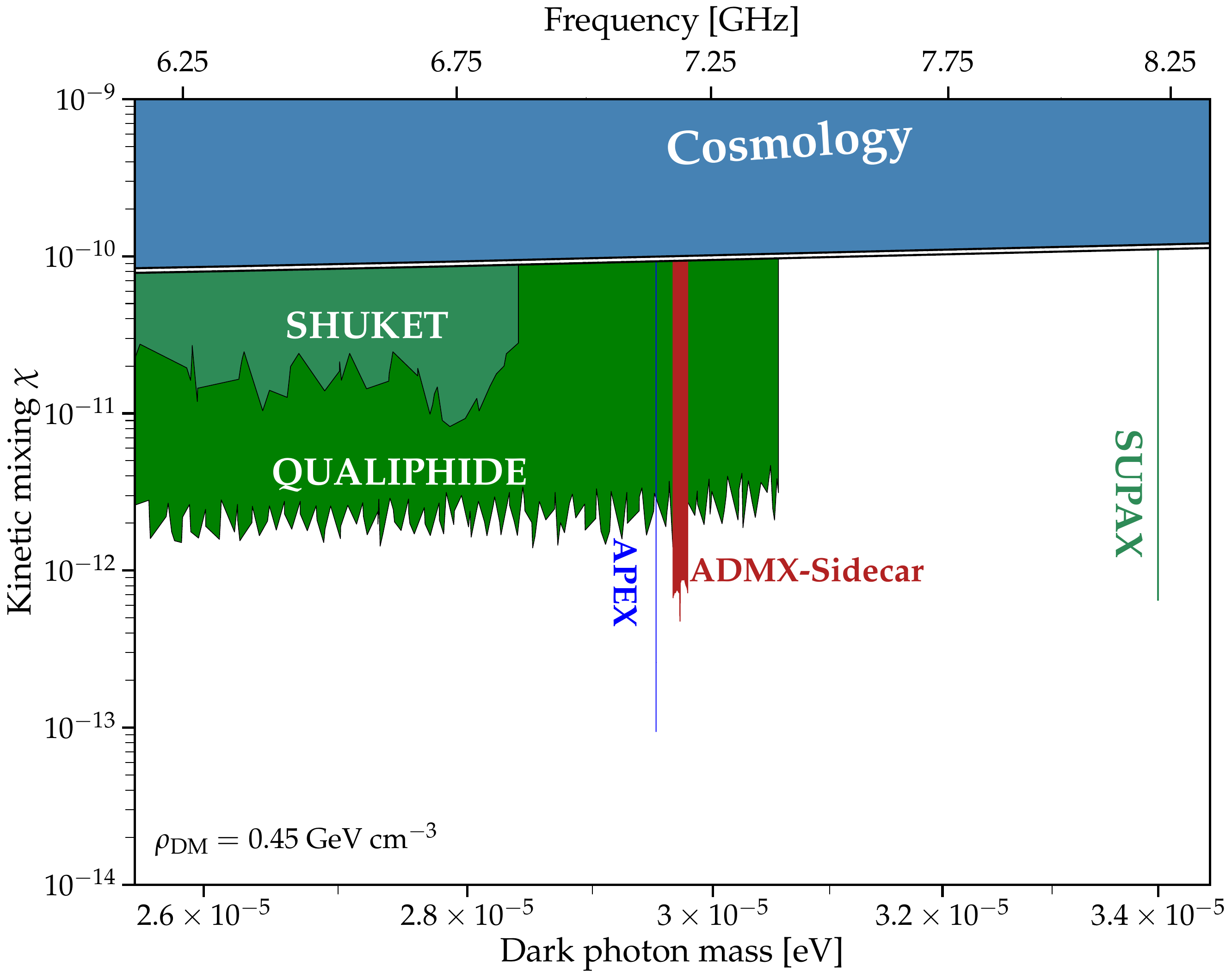}
	\caption{Final dark photon constraint, shown both in close up (left) and within the wider context of other dark photon constraints (right).
    We constrain the kinetic mixing parameter $|\chi|<3.7\times10^{-13}$ around $m_A \simeq 29.5 $  $\mu$eV ($7.139$ GHz) (90 \% confidence level).
    Other constraints shown are from Refs.~\cite{ADMX:2018ogs, McDermott:2019lch, Brun:2019kak, Ramanathan:2022egk, Schneemann:2023bqc}.
    Figure production utilizes the AxionLimits code \cite{AxionLimits}.
    }
	\label{fig: final constraint}
\end{figure*}

This being the case, we can place limits on the contribution of dark photon dark matter to the observed signal power.
For each mass under consideration, we construct reference spectra with $\chi=1$ and the assumed form of the lab-frame DM frequency distribution 
\begin{align}
    F(f) \simeq 2&\left(\frac{f - f_{DM}}{\pi}\right)^{1/2}\left(\frac{3}{1.7 f_{DM}v_{DM}^2}\right)^{3/2}\\\nonumber
    &\times\exp\left(-\frac{3(f - f_{DM})}{1.7f_{DM}v_{DM}^2}\right)\,,
\end{align}
which satisfies $\int\,df F(f) = 1$, where $f_{DM} = m_A/2\pi$ and $v_{DM}\simeq 9\times10^{-4} c$ is the DM virial velocity \cite{Cervantes:2022epl}.
Convolution of the dark photon signal power in \eqref{eq: power} with this lineshape then yields the reference power $P_{\rm ref}$ in each bin, with the corresponding likelihood given (up to an irrelevant overall normalization factor) in terms of the product of different bins via 
\begin{equation}
    p(P_e | m_A, \chi) = \prod_i \frac{1}{\sqrt{2\pi\sigma_P^2}}\exp\left(-\frac{(P_e - P_{\rm ref}\chi^2)^2}{2\sigma_P^2}\right)\,,
\end{equation}
where the factor of $\chi^2$ corrects for our original choice of $\chi = 1$. 
In principle the product runs over all frequency bins, however due to the finite DM linewidth we use only the 10 bins beginning at $f_{DM}$.
From here we then find the 90 \% confidence limit $|\chi|<3.3\times 10^{-13}$ shown in Fig.~\ref{fig: final constraint}.

The uncertainty in our analysis is primarily statistical in nature, driven by the 1.7\,\% relative uncertainty in $P_e$, whilst systematic uncertainties in the experimental parameters such as $\beta$ (relative uncertainty 0.7\,\%) and $Q_L$ (relative uncertainty 0.3\,\%) are subleading.

We can compare this result with theoretical expectation derived from the Dicke radiometer equation ${\rm SNR} = P_s/P_n\times\sqrt{b\, t_{\rm int}}$, although care is required here in that this equation assumes all of the signal power falls within the bandwidth $b$, whilst in our case $b$ for this pathfinder data taking run is chosen to be much smaller than the assumed DM linewidth.
Correcting for this modification to $P_s$ and rearranging we find
\begin{equation}
    \chi \simeq \sqrt{\frac{\beta + 1}{\beta}\frac{{\rm SNR}\,T_n}{\eta m_A\rho_A V_{\rm eff} Q_L}\left(\frac{\Delta_A}{b}\right)}\left(\frac{b}{t_{\rm int}}\right)^{1/4}\,,
\end{equation}
where $T_n \simeq 5.5$ K is the noise temperature and $\Delta_A\simeq m_A v_{DM}^2/(2\pi)$ is the DM linewidth.
Using the values in Table~\ref{tab: experimental parameters} we find $\chi \simeq 4.8 \times 10^{-13}$ in approximate agreement with our derived limit.

\textbf{Conclusions/discussion.}
The dark photon is a promising candidate for the DM which comprises most of the matter in our visible Universe.
It is also an attractive experimental target, since via kinetic mixing with the Standard Model it can resonantly convert to photons in an electromagnetic cavity.

We have performed a search for this type of DM using a pathfinder cavity experiment, representing the first stage of the APEX (Axion and dark Photon EXperiment) program. 
Finding no statistically significant excess, we constrain the dark photon kinetic mixing parameter $|\chi|<3.7\times 10^{-13}$ around $m_A \simeq 29.5$  eV ($7.139$ GHz) (90 \% confidence level).
This result exceeds other constraints on dark photon dark matter in this frequency range by roughly an order of magnitude.

Having established the initial experimental operation, data taking and analysis, going forward we plan to implement mechanical scanning of the cavity to greatly increase the available frequency range, in line with the approach taken by other haloscope experiments.
The future addition of a magnetic field will also enable this experiment to also search for axion DM, via the Primakoff effect. Methods to improve detection sensitivity, such as the dual-path interferometry scheme explored in Ref.~\cite{Yang:2022uil}, are also of particular interest.

\medskip
\textbf{Acknowledgements.} 
This work is supported in part by the Scientific Instrument Developing Project of the Chinese Academy of Sciences (YJKYYQ20190049), the International Partnership Program of Chinese Academy of Sciences for Grand Challenges (112311KYSB20210012), the National Natural Science Foundation of China (11875062, 11947302, 11905149, 12047503, 12074117, 12150010, 61833010, 12061131011, 12150410317, 12275333 and 12375065), the Key Research Program of the Chinese Academy of Sciences (XDPB15), the Beijing Natural Science Foundation (IS23025) and the Natural Science Basic Research Program of Shaanxi (2024JC-YBMS-039).

\end{document}